\documentclass[aps,twocolumn]{revtex4}
\usepackage[latin1]{inputenc}
\usepackage{graphicx}

\begin{document}

\title{Electronic and magnetic properties of a hexanuclear ferric wheel}
\author{H. Nieber}
\email{h.nieber@tu-bs.de}
\author{K. Doll}
\author{G. Zwicknagl}
\affiliation{Institut f\"ur Mathematische Physik, TU Braunschweig,
Mendelssohnstra{\ss}e 3, D-38106 Braunschweig}

\pacs{ }

\begin{abstract}
The electronic and magnetic properties of the hexanuclear ferric
wheel [LiFe$_6$(OCH$_3$)$_{12}$-(dbm)$_6$]PF$_6$ have been studied with
all-electron Hartree-Fock and full-potential density functional
calculations. The best agreement for the magnetic exchange coupling
is at the level of the B3LYP hybrid functional. Surprisingly,
the Hartree-Fock approximation gives the wrong sign for the exchange
coupling. The local density
approximation and the gradient corrected functional PBE strongly overestimate
the exchange coupling due to the too large delocalization of the
$d$-orbitals. These findings are supported by results from the
Mulliken population analysis
for the magnetic moments and the charge on the individual atoms.
\end{abstract}
\maketitle

\section{Introduction}

Molecular magnetism has received great attention in the past few years,
especially after
a series of molecules has been discovered which show magnetism due to
a molecular origin (see the review articles\cite{caneschi1999,pilawa1999}).
One particular type
of molecular magnets are the so-called ferric wheels where iron atoms are
arranged ring-like. So far, wheels with 6 up to 18 iron atoms have
been synthesized\cite{caneschi1999}.
This has also triggered interest among theorists. The description
of magnetism still poses challenges, and only in the last years
ab-initio calculations of large molecular magnets have become feasible
\cite{postnikovpsik}.

The early studies employed wave-function based methods: for example,
the exchange interaction in transition metal oxides was investigated
at the level of periodic Hartree-Fock calculations \cite{Towler1994}.
Usually, it is found that the Hartree-Fock approximation strongly
underestimates the exchange coupling
(e.g. \cite{Towler1994,ricart1995,catti1995}).
This was also demonstrated
for smaller molecules, where demanding  configuration interaction
calculations were feasible: it was shown, that for a proper description
of the magnetic interactions, highly correlated wave functions are
necessary\cite{Fink1,Fink2}.

With the help of an appropriate embedding,
the bulk could be approximatively described with clusters.
Then, wave-function based correlation methods could be applied and
a value in reasonable agreement with
the experimental value of the exchange coupling was deduced
\cite{degraaf,moedl}.

In the past few years, density functional calculations have become
more and more widespread, and have also been applied to molecular
magnets (e.g. \cite{postnikovpsik,ruiz2003,zeng1}).
They have the advantage of including approximatively the effects of
electronic correlation, at a very low computational expense. However,
it is often observed that exchange couplings, computed at the
density functional level (especially the local density approximation),
strongly overestimate the experimental value. In a comparative study,
it was shown that hybrid functionals, which include the exact (Fock)
exchange, improve this and
provide surprisingly accurate values \cite{iberio}. This was
also observed earlier
in cluster calculations which were used to model the solid
\cite{MartinIllas1997}.

The ferric wheel with six iron atoms is one of the smaller molecular magnets
and thus ab-initio calculations can be performed at a relatively low expense,
besides the approach using model Hamiltonians \cite{Normand,Honecker}.
Various molecules have been synthesized
\cite{Caneschi1995angew,Caneschi1996,Abbati1997,Lascialfari1997,Pilawa1997,Saalfrank1997,Waldmann1999,Affronte1999,Cornia1999,Affronte2002,Pilawa2003} and
techniques such as nuclear magnetic resonance, high-field
DC and pulsed-field differential magnetization experiments, susceptibility
measurements, high field torque magnetometry, inelastic neutron scattering
and heat capacity measurements have been applied.

The experimental values for the exchange couplings,
extracted from susceptibility measurements
or from the singlet-triplet gap, are typically around $\sim$20 K
\cite{caneschi1999,Waldmann1999,Cornia1999,Affronte1999,Pilawa2003}, although also larger values have been suggested (e. g. 38 K\cite{Lascialfari1997}).
As there
are still only few ab-initio calculations\cite{postnikov2003,postnikov2004},
these systems are therefore
an interesting object to study.

This article will deal with one of the six-membered iron ferric wheels
\cite{Abbati1997}. In previous studies on a similar system with a gradient
corrected functional, an overestimation of the calculated exchange coupling
by a factor of $\sim$ 4 was found and it was argued that schemes
such as LDA+$U$ or self-interaction correction might
resolve this problem\cite{postnikov2003,postnikov2004}.
As was argued above, hybrid functionals might
be a different way to deal with this problem. We therefore studied this
ring at the level of various functionals including hybrid functionals,
and the traditional Hartree-Fock approach.

\section{Method}
\label{methodsection}

All the calculations were done with the code CRYSTAL2003 \cite{crystal,dovesi}.
We employed the Hartree-Fock (HF) method (in the variant of unrestricted
Hartree-Fock),
the local density approximation (LDA) with the correlation functional of
Perdew and Zunger\cite{PZ},
the gradient corrected functional of Perdew, Burke and
Ernzerhof (PBE)\cite{PBE}
and the hybrid functional B3LYP (a functional with admixtures,
amongst others, of functionals by Becke, Lee, Yang and Parr).
As described later on in this section, we performed the calculations
mainly on a molecular system. Therefore, we also used a basis set
library for molecular basis sets\cite{Basistext} rather then the
CRYSTAL basis set library which contains essentially basis sets
for periodic systems.
We chose a [\textit{5s4p2d}] basis set
for iron \cite{Fe}, where the diffuse exponent 0.041148 was omitted,
a [\textit{2s}] basis set for hydrogen\cite{ditchfield},
a [\textit{3s2p}] basis set for carbon\cite{hehre} where the exponent
0.1687144 was replaced with 0.25 to avoid convergence problems,
a [\textit{3s2p}] basis set for oxygen\cite{oxy} with an additional
$d$-exponent of 0.8 and a [\textit{3s2p}] basis set for lithium \cite{oxy}
where the exponent 0.2 was omitted. After the mentioned modifications,
the final basis sets were thus of the size
[\textit{4s3p2d}] (iron), [\textit{3s2p}] (carbon), [\textit{3s2p1d}]
(oxygen), [\textit{2s}] (hydrogen) and [\textit{2s1p}] (lithium).

To explore the influence of further basis functions, various tests were
performed: calculations with an additional $sp$-shell (exponent 0.04) and an
additional $d$-shell (exponent 0.2) at the iron atom have been carried out,
at the B3LYP and HF level.
Also calculations with an additional $sp$-shell (exponent 0.12) at the oxygen atoms were
performed at the B3LYP level.
This diffuse exponent led to linear dependence problems at
the HF level. However, a HF calculation where the outermost exponent 0.270006
for oxygen was replaced with two exponents (0.5, 0.2) was numerically stable.
No significant changes for the computed exchange couplings
were observed in any of these aforementioned tests.
Therefore the basis sets, as described in the first paragraph of this
section, can be considered reliable, and all the results were obtained with
these basis sets, if not explicitly stated otherwise.

The properties
of the ferromagnetic (FM) state (all spins parallel, total spin 30 $\mu_B$)
and of the antiferromagnetic (AF) state (spins alternating up and down,
total spin 0) were computed with each method.
To obtain the local magnetic moments, a Mulliken population analysis was
performed.

The calculations were performed on an isolated
molecule where the full symmetry of the molecule was exploited
(ferromagnet: $C_{3i}$, antiferromagnet: $C_{3}$).
The geometry was chosen according to the measurements by
Abbati et al \cite{Abbati1997} and is displayed in figure \ref{geometry}.
The isolated molecule was charged with +1, as lithium
is ionized and charge neutrality is
restored by a PF$_6^-$ ion for this particular ferric wheel.
To assess the influence of the centered lithium ion it was also
removed in some of the calculations.

\section{Results}
\label{resultssection}
The total energies, the difference in total energies of the ferromagnet and the
antiferromagnet and the exchange parameters $J$
are given in table \ref{deltaenergy}.
To obtain the exchange parameters \textit{J} from the calculation, the spin Hamiltonian
of the Ising model was used:

\begin{displaymath}
\textbf{H} = -J\sum_{i=1}^5 S_i \cdot S_{i+1}-J\cdot S_6\cdot S_1
\end{displaymath}

The variable $i$ corresponds to the site index of the iron atoms, which
in this model possess a spin of $S=5/2$.
With this Hamiltonian, the energy difference between ferromagnet(FM) and antiferromagnet(AF) then
corresponds to:

\begin{displaymath}
E_{tot}^{FM}-E_{tot}^{AF}=-12JS^2
\end{displaymath}

Therefore, the exchange parameter \textit{J} results from
the difference of large numbers.
Still the results are numerically stable, as can be seen in table \ref{deltaenergy}.
For the particular ferric wheel under consideration,
the experimental value for the exchange coupling
was determined to be -21 K \cite{Abbati1997,caneschi1999}.

This can now compared with the computed values.
One interesting result is the ferromagnetic exchange parameter obtained at the
Hartree-Fock level (+7 K ) which
does not reproduce the real antiferromagnetic nature
of the molecule. This
result was stable $(\pm2 K)$ with respect to all possible
variations in the basis set as mentioned in section \ref{methodsection},
and also when the central lithium ion was removed.

Such a problem did not show up with any of the density functionals.
The magnitude of $J$ of the B3LYP calculation (-31 K)
is about 50$\%$ larger than the experimental value.
In the PBE and LDA calculations, it is even overestimated
by a factor of $\sim$ 5 (PBE: -108 K, LDA: -120 K).
This agrees well with an earlier calculation
for a similar ferric wheel by Postnikov et al \cite{postnikov2003,postnikov2004} with the PBE functional, where $J$ was also
largely  overestimated (-80 K). The authors argued,
that the $d$-orbitals in density functional calculations with functionals
such as LDA or PBE, were not sufficiently localized
to compare with experiment, and therefore
the resulting exchange parameters would overestimate
the experimental values. They therefore
suggested to use methods such as LDA+$U$ or self-interaction corrections
to improve this shortcoming. The B3LYP functional
may be viewed as another possibility
to remedy this problem, as it interpolates between the pure Hartree-Fock
approach (with an unscreened Coulomb interaction and the correct Fock
exchange, but without electronic correlation), and standard functionals
(including electronic correlation, but too delocalized $d$-orbitals).
This was already demonstrated for bulk NiO \cite{iberio},
where the B3LYP value
for the exchange couplings was also found to be about 50 \% larger
than the experimental value, and LDA overestimated also by a factor
of $>4$. Also, the total density of states was calculated at the HF and B3LYP level and is shown in figure \ref{doss}.
Note that these calculations were carried out as bulk calculations.
The HOMO-LUMO gap at the HF level is about ten times larger than
at the B3LYP level, but there is no significant difference between the
ferromagnetic and the antiferromagnetic state at one level of theory.
However, the smaller gap will favor hybridization and thus the more delocalized situation.

Comparing the results, the wrong sign for \textit{J} at the HF level is rather surprising as it was
usually observed that the Hartree-Fock result underestimated the experimental
value, but with the correct sign, with few exceptions\cite{ruiz2003}.
A possible source for the discrepancy can be found in the geometry of the
molecule. For a system where the magnetic centers are in line with the bonding bridge atom,
consideration of correlation leads to a strong increase of the antiferromagnetic exchange parameter
as it was found for NiO by de Graaf et al \cite{degraaf} and for other complexes by Fink et al\cite{Fink1}.
In our case, the Fe-O-Fe bonding angle is in the range of 100 degrees\cite{Abbati1997}
and for such systems (and similarly for other ferric wheels, see e.g. Waldmann et al\cite{waldmann2001})
the exchange parameter is supposed to be less antiferromagnetic or even
ferromagnetic compared to in-line geometries which is a result of the ordering of the magnetic orbitals
according to the Goodenough-Kanamori rules\cite{kahn}.
In addition, due to the to large HOMO-LUMO gap,
the system is too ionic at the HF level
and thus the overlap of the orbitals is underestimated,
which also has an impact on the magnitude of the computed exchange interaction.
Thus the small antiferromagnetic coupling and
the underestimation of the exchange parameter due to the neglect of correlation
at the HF level may be the reason for the wrong sign.

The central lithium atom has an electrostatic influence to the molecule
and the lithium ionic radius affects the diameter of the ferric wheel.
Therefore, when the lithium is removed and the geometry input is not changed
as in our case, no impact on the exchange parameter is expected.

To analyze the magnetic states, the magnetic moments were computed.
They are displayed in table \ref{magneticmoments}.
A decrease of the local magnetic moment of the iron atoms from HF over B3LYP
and PBE to LDA is observable, where the moment becomes more delocalized and
is transferred to the surrounding oxygen atoms. This is particularly
obvious in the ferromagnetic state. In the antiferromagnetic state, the iron
magnetic moments are virtually the same as in the ferromagnetic state,
apart from the sign.
However,
the bridge oxygens carry virtually no magnetic moment, as there is an overlap
of up and down spin density at these sites,
originating from two neighboring irons, which as a whole cancels. For the same reason
there is no magnetic moment at the centered lithium,
and even in the ferromagnetic case the lithium spin is virtually zero.

This feature is reflected in the spin-density plots in figure
\ref{spin_merge} and in larger resolution in figure \ref{rescaled}. For a more detailed study,
an one-dimensional plot of the spin-density at the LDA level
along a line from O(apical) via Fe, O(bridge) and Fe to O(apical)
was performed (figure \ref{1Dspindensity}).
The iron atoms subsist in a high-spin state
$3d^ 5_{\uparrow}\textit{d}^ 0_{\downarrow}$,
but except for the HF method the moment is
closer to $4\mu_B$ as it was also obtained by Postnikov et al for a slightly
different molecule\cite{postnikov2003,postnikov2004}.
As can be seen in table \ref{magneticmoments}, the local magnetic moment of the apical
oxygen atoms increases
from HF over B3LYP and PBE to LDA. In addition,
there is a local magnetic moment at
the carbon C(1) atom (see figure \ref{geometry} and table \ref{magneticmoments})
which is about 4 times (HF)
to 2 times (LDA) smaller than the local magnetic moment of the apical
oxygen atom. Corresponding to the moment of the C(1) atom there is a small
magnetic moment at the carbon atoms C(2) and C(3), but with opposite sign and
a much smaller value. Further on there is actually a moment at the first
carbon atom (next to the carbon atoms C(2) and C(3)) of each $C_6H_5$ ring. Its value is of the same
order as the value of the C(2) and C(3) carbon atoms moments,
but with opposite sign.
In the ferromagnetic state, there is a small magnetic moment at the C(4) bridge carbon
atoms corresponding to the magnetic moment of the bridge oxygen atoms which is also, as
already mentioned above, increasing from HF over B3LYP and PBE to LDA.

The results from the Mulliken population analysis
of the iron, oxygen (bridge and apical), lithium and carbon
atoms are given in tables \ref{fepop} and \ref{Charge}.
The net charge of iron is between 1.27 and 2.16,
at the various levels of theory, and thus far away from a formal
charge\cite{Abbati1997} of +3. Again, Hartree-Fock gives a very localized
picture with the largest net charge (2.16), and LDA a delocalized
picture with an iron charge of only 1.27. Oxygen carries essentially a
single negative charge ($\sim$ -0.8 in LDA, $\sim$ -1.1 in HF),
the carbon atoms which have oxygen as a neighbor
all have donated charge (0.4 - 0.8 $|e|$, depending on the site and
the level of theory).
The lithium charge is between 0.4 (LDA) and 0.8 (HF).
A plot of the difference of the charge density was performed between the values at the LDA and HF level
(figure \ref{chargedifference}). There is a positive charge density difference at the iron atoms
because of the smaller charge at the LDA level compared to HF,
the results for the other atoms follow from tables \ref{fepop} and \ref{Charge}.
This plot thus visualizes the more covalent picture obtained at the LDA level.
The stronger covalency also becomes obvious in the Mulliken overlap population:
for example, the overlap population between Fe and the neighbouring oxygens atoms
is 0.11-0.12 at the LDA level, but only 0.07-0.08 at the HF level.

\section{Summary}
\label{summarysection}

\textit{Ab-initio} calculations for a hexanuclear ferric wheel
\cite{Abbati1997} were performed.
The exchange coupling parameter $J$  was computed
at various levels of theory, with the B3LYP functional providing the
best agreement with experiment (-31 K, experiment: -21 K). LDA and PBE
were found to grossly overestimate the exchange coupling parameter
due to the too large delocalization
of the $d$-orbitals. Surprisingly, the calculation at the Hartree-Fock level
led to a ferromagnetic coupling. The electronic
population and the spin densities were calculated. The total spin
is distributed over several sites, besides the iron atom all of
the neighboring oxygens carry some spin. The delocalization increases from
HF over  B3LYP, PBE to LDA which shows the strongest delocalization and
thus the smallest magnetic moment on the iron site.

\section{Acknowledgments}

Most of the calculations were performed at the compute-server \textit{cfgauss}
(Compaq ES 45) of the data processing center of the TU Braunschweig.
The geometry plot of the molecule was performed with MOLDEN\cite{molden}.

\begin{widetext}
\newpage
\begin{table}
\begin{center}
\caption{\label{deltaenergy}
Total energies ($E_h\equiv$ Hartree), the differences in total energies for the ferromagnetic
(FM) and the antiferromagnetic (AF) state and exchange parameters \textit{J} }
\vspace{5mm}
\begin{tabular}{ccccc}
method &  FM total energy \hfill     $(E_h)$& AF total energy $(E_h)$\hfill & difference of total energy \hfill $(mE_h)$& exchange parameter $J$ \hfill (K) \\ \hline
HF & -13295.7328681 & -13295.7312480 & -1.62 & +7 \\
HF (without Li) & -13287.8311111 & -13287.8293232 & -1.79 & +8 \\
B3LYP & -13334.5067552 &  -13334.5140876 & 7.33 & -31 \\
PBE & -13329.9510763 & -13329.9768112 & 25.74 & -108\\
LDA & -13273.9493819 & -13273.9779682 & 28.58 & -120\\
\end{tabular}
\end{center}
\begin{center}
\caption{\label{magneticmoments}
Local magnetic moments at various sites, in $\mu_B$, for the ferromagnetic
(FM) and the antiferromagnetic (AF) state }
\vspace{5mm}
\begin{tabular}{p{1.5cm}p{1cm}p{1cm}p{1cm}p{1cm}p{1cm}p{1cm}p{1cm}p{1cm}p{1cm}p{1cm}p{1cm}p{1cm}}
\centering method & \multicolumn{2}{c} {Fe }& \multicolumn{2}{c}  {O (apical)} &\multicolumn{2}{c}  {O (bridge)}
& \multicolumn{2}{c} {C(1)}& \multicolumn{2}{c}  {C(2)/C(3)} & \multicolumn{2}{c} {C(4)}\\
\centering & \centering AF &\centering  FM &\centering AF &\centering FM &\centering  AF &\centering FM
& \centering AF & \centering FM & \centering AF & \centering FM & \centering AF &  \multicolumn{1}{c}{FM}\\ \hline
\centering HF & \centering $\pm4.79$ & \centering $+4.79$ & \centering $\pm0.04$ & \centering $+0.04$ & \centering$\approx 0$ &\centering$+0.07$
& \centering $\pm0.008$ & \centering $+0.009$ & \centering $\mp0.006$ & \centering $-0.006$ &  \centering$\approx 0$ & \multicolumn{1}{c}{$\approx 0$}\\
\centering B3LYP & \centering $\pm4.32$ & \centering $+4.34$ & \centering $\pm0.10$ & \centering $+0.10$ & \centering$\approx 0$ &\centering$+0.20$
& \centering $\pm0.031$ &  \centering $+0.032$ & \centering $\mp0.007$ & \centering $-0.007$ & \centering$\approx 0$ & \multicolumn{1}{c}{+0.005}\\
\centering PBE & \centering $\pm4.10$ & \centering $+4.19$ & \centering $\pm0.10$ & \centering $+0.11$ & \centering$\approx 0$ &\centering$+0.22$
& \centering $\pm0.041$ &  \centering $+0.046$ & \centering $\mp0.005$ & \centering $-0.003$ & \centering$\approx 0$ & \multicolumn{1}{c}{+0.009}\\
\centering LDA & \centering $\pm3.96$ & \centering $+3.94$ & \centering $\pm0.11$ & \centering $+0.11$ & \centering$\approx 0$ &\centering$+0.20$
& \centering $\pm0.040$ & \centering $+0.047$ & \centering $\mp0.003$ & \centering $-0.006$ & \centering$\approx 0$ & \multicolumn{1}{c}{+0.010}\\
\end{tabular}
\end{center}
\begin{center}
\caption{\label{fepop}
Mulliken charge of Fe, in $|e|$. Note that the charge is virtually
identical for the ferromagnetic and the antiferromagnetic state.}
\vspace{5mm}
\begin{tabular}{ccccc}
method & net charge \hfill  &   \textit{s} \hfill & \textit{p} \hfill  & \textit{d} \\ \hline
HF & 2.16 & 6.28 & 12.28 & 5.29 \\
B3LYP & 1.56 & 6.38 & 12.37 & 5.69 \\
PBE & 1.40 & 6.39 & 12.37 & 5.84 \\
LDA & 1.27 & 6.43 & 12.41 & 5.89 \\
\end{tabular}
\end{center}
\begin{center}
\caption{\label{Charge} Charge at various sites, in $|e|$.
Note that the charge is virtually
identical for the ferromagnetic and the antiferromagnetic state.}
\vspace{5mm}
\begin{tabular}{ccccccc}
method & Li \hfill&  O (apical) \hfill     & O (bridge) \hfill & C(1) \hfill & C(2)/C(3) \hfill & C(4) \hfill \\ \hline
HF & 0.77 & -1.10 & -1.18 & -0.22 & 0.82 & 0.50 \\
B3LYP & 0.53 & -0.86 & -0.88 & -0.11 & 0.62 & 0.45\\
PBE & 0.47 & -0.79 & -0.80 & -0.10 & 0.56 & 0.38 \\
LDA & 0.38 & -0.76 & -0.75 & -0.09 & 0.56 & 0.36 \\
\end{tabular}
\end{center}
\end{table}

\newpage
\begin{figure}[p]
\caption{Geometry of the ferric wheel.
Hydrogen atoms have been omitted for clarity.}
\label{geometry}
\center\includegraphics[width=13cm,angle=270]{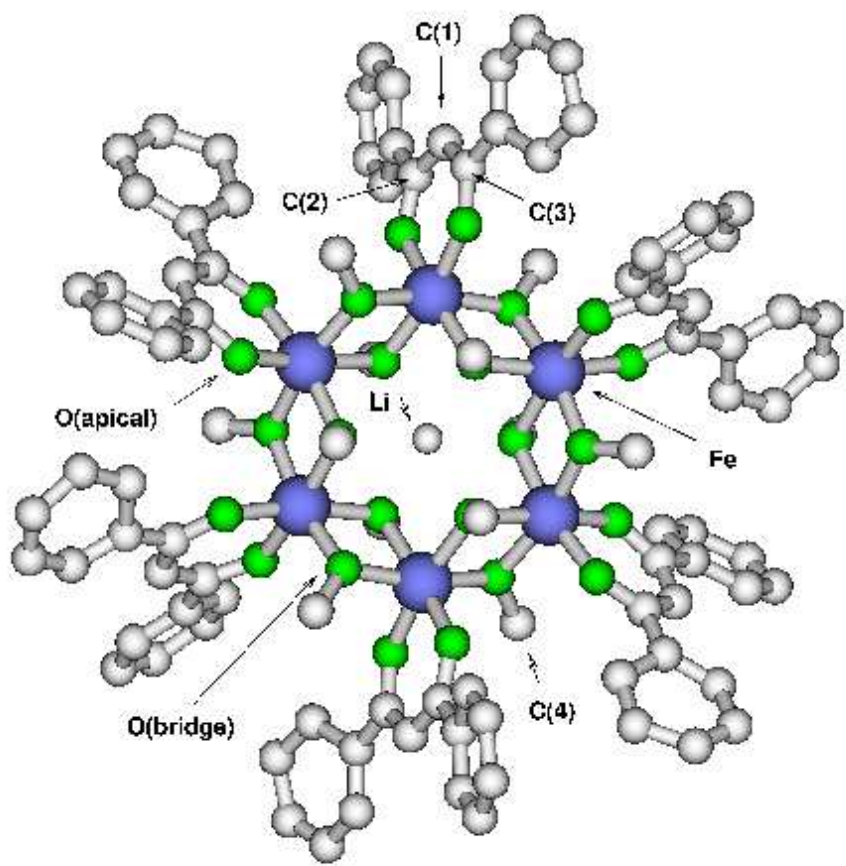}
\end{figure}
\newpage
\begin{figure}[p]
\caption{The total density of states for the ferromagnetic and antiferromagnetic state at the B3LYP level (upper two graphs) and the HF level (lower two graphs). For this purpose, the calculations had to be carried out as bulk calculations,
with a lattice constant of 100 {\AA} to avoid interactions between the molecules,
and with the central lithium ion removed in order to have
a non-charged unit cell.
The top of the valence band is at -0.12 E$_h$ (B3LYP) and at -0.26 E$_h$ (HF).}
\label{doss}
\centerline{\includegraphics[width=15.5cm,angle=270]{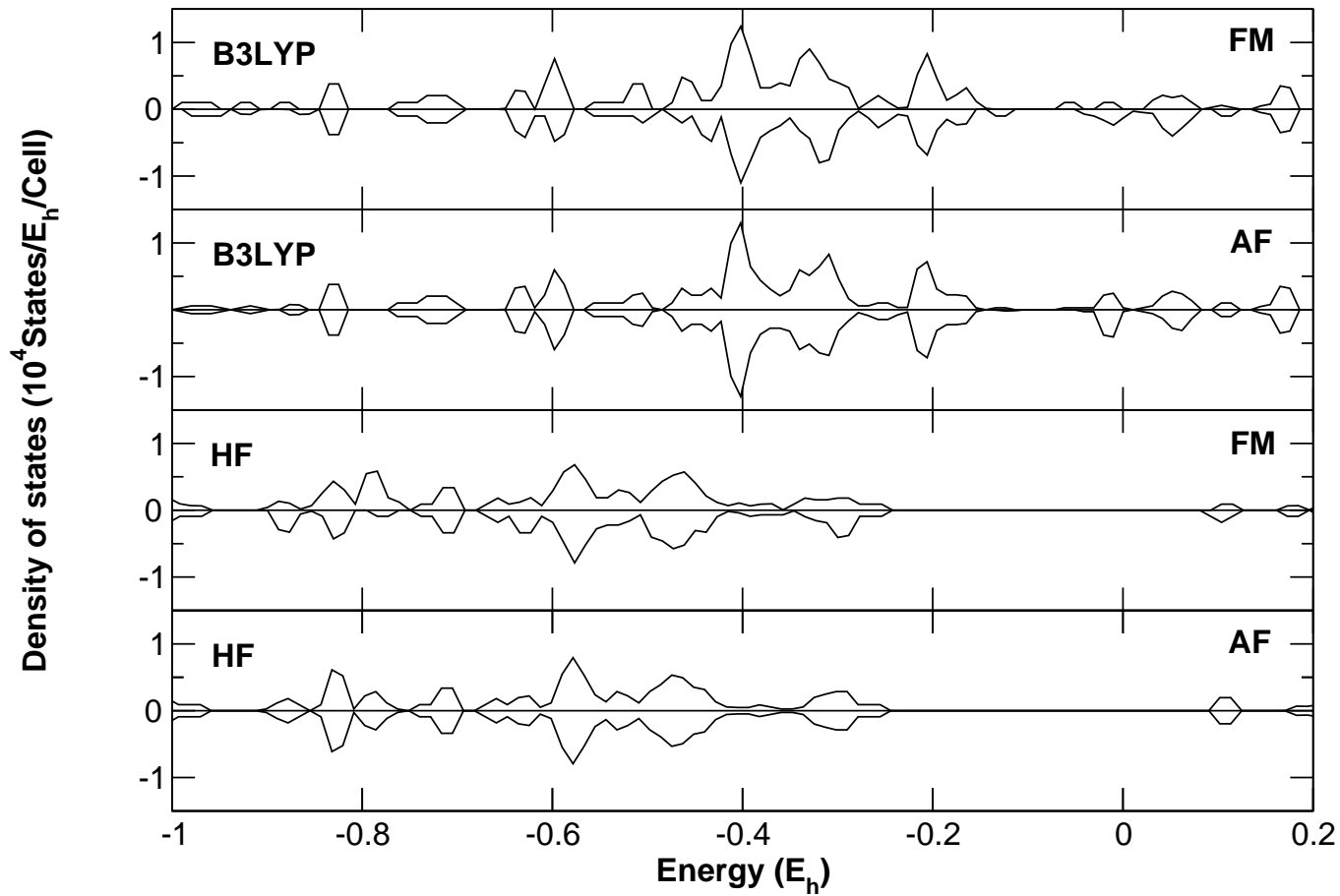}}
\end{figure}

\newpage
\begin{figure}
\caption{Spin densities at the LDA level for the ferromagnetic (FM) and the antiferromagnetic (AF) state.
For all figures, the contour lines range from -0.0004 to 0.0005 in steps of 0.000035 electrons$/($a.u.$)^3$.
Full lines indicate positive spin density and dashed lines indicate negative spin density.}
\label{spin_merge}
\centerline{\includegraphics[width=13cm,angle=0]{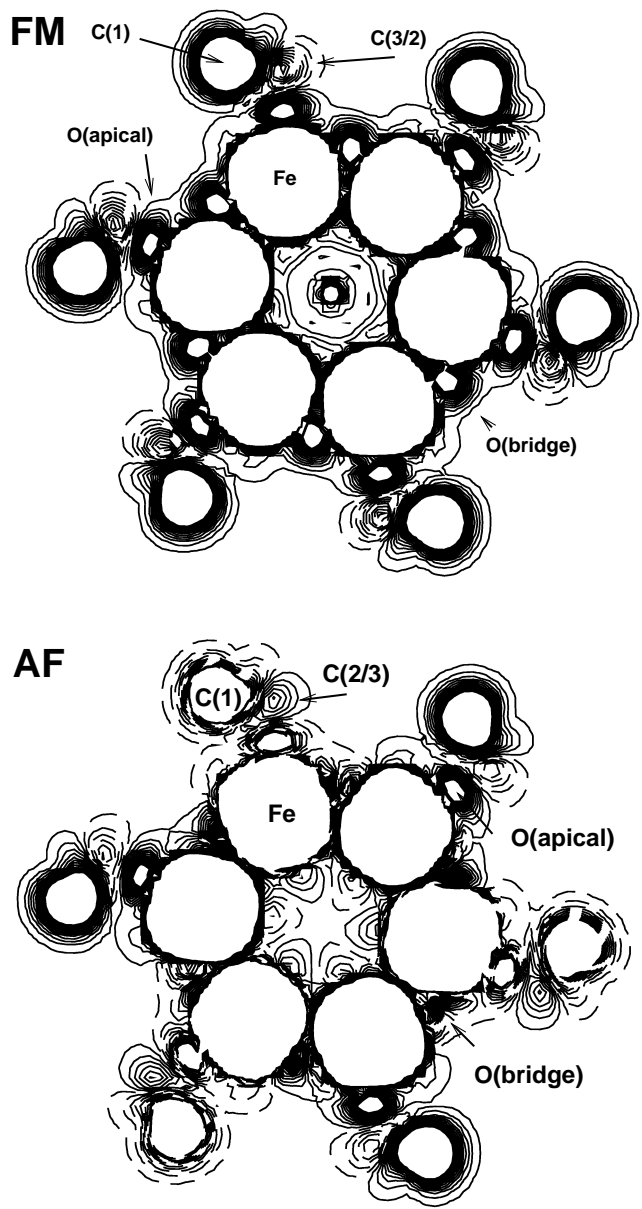}}
\end{figure}

\newpage
\begin{figure}
\caption{Rescaled plots of the spin densities for the ferromagnetic (FM) state at the LDA level (first graph) and
for the antiferromagnetic (AF) state at the LDA (second graph) and the HF level (third graph).}
\label{rescaled}
\centerline{\includegraphics[width=13cm,angle=0]{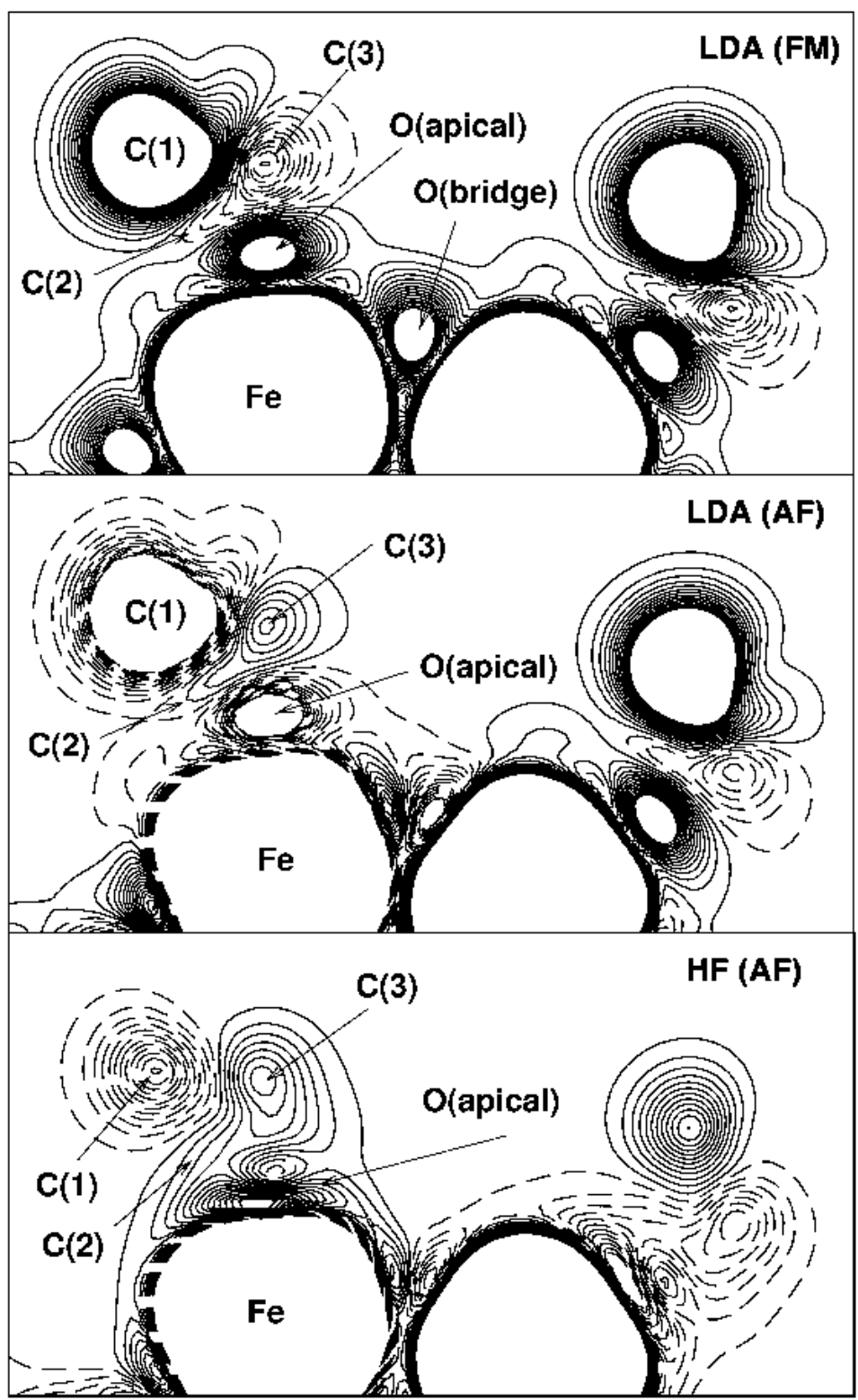}}
\end{figure}

\begin{figure}
\caption{One-dimensional plot of the spin density at the LDA level for the antiferromagnetic state along a line between certain atoms of the molecule, see labeling of the abscissa and insert caption. The scaling of the ordinate refers to the values in table \ref{magneticmoments}. }
\label{1Dspindensity}
\centerline{\includegraphics[width=15.5cm,angle=0]{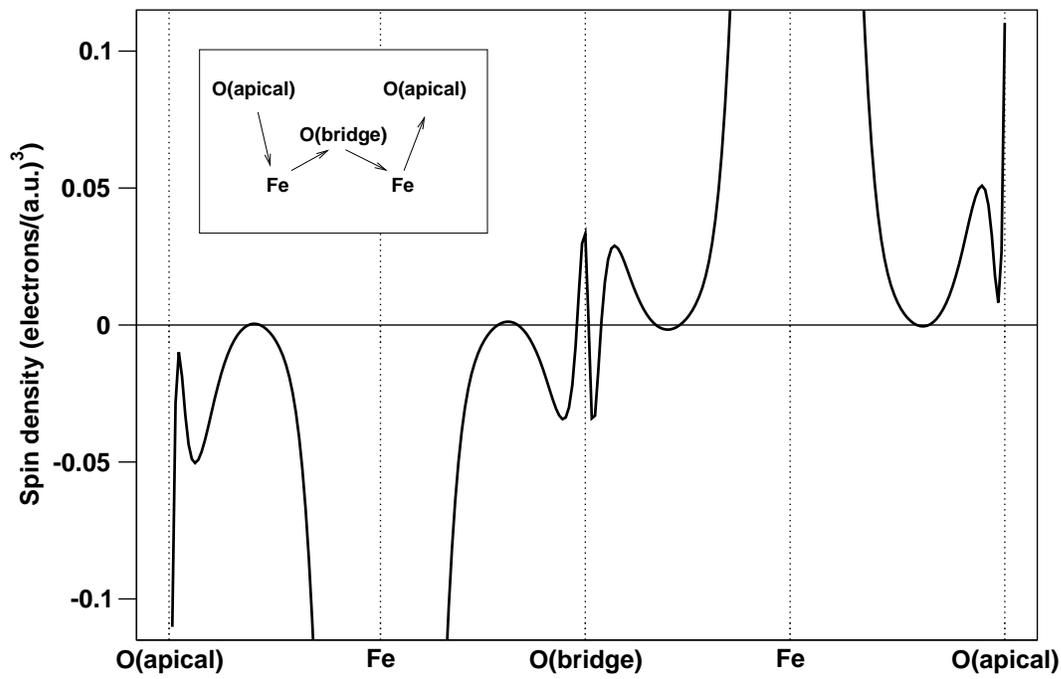}}
\end{figure}

\begin{figure}
\caption{Rescaled plot of the charge density difference between the LDA and the HF level for the antiferromagnetic state. The contour lines range from -0.005 to 0.005 in steps of 0.0007 electrons$/($a.u.$)^3$. Full lines indicate positive charge density difference and dashed lines indicate negative charge density difference.}
\label{chargedifference}
\centerline{\includegraphics[width=13cm,angle=0]{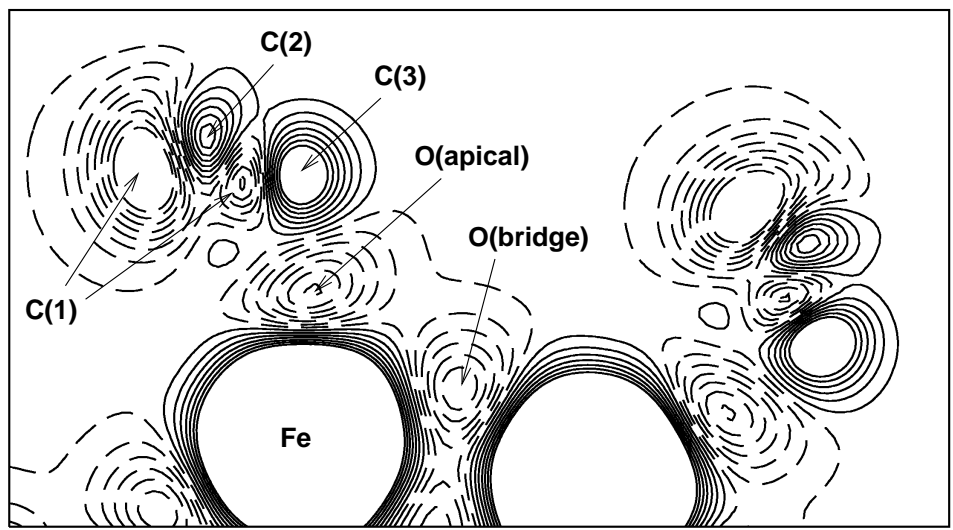}}
\end{figure}

\end{widetext}

\end{document}